\documentclass[twocolumn]{aastex63}

\usepackage{graphicx}
\usepackage{ulem}
\usepackage{amsmath}
\usepackage{wrapfig}
\usepackage[thinlines]{easytable}
\usepackage{tikz}
\usepackage{hyperref}

\newcommand{\be}{\begin{eqnarray}}
\newcommand{\ee}{\end{eqnarray}}
\newcommand{\lp}{\left(}
\newcommand{\rp}{\right)}
\newcommand{\lb}{\left[}
\newcommand{\rb}{\right]}

\graphicspath{{./}{figures/}}

\begin{document}
\shorttitle{}
\shortauthors{Piro \& Nakar}

\title{Early Emission from Double Detonation Type Ia Supernovae}

\correspondingauthor{Anthony L.\ Piro}
\email{piro@carnegiescience.edu}

\author[0000-0001-6806-0673]{Anthony L.\ Piro}
\affiliation{The Observatories of the Carnegie Institution for Science, Pasadena, CA 91101, USA}

\author[0000-0002-4534-7089]{Ehud Nakar}
\affiliation{School of Physics \& Astronomy, Tel Aviv University, Tel Aviv 69978, Israel}

\begin{abstract}
A popular model for Type Ia supernovae (SNe Ia) is the detonation of a CO white dwarf (WD) that is triggered by the prior detonation of a thin surface layer of helium, known as a double detonation (DD). We explore the unique early electromagnetic signatures that are expected from collision of the CO detonation with the He detonation. The three features are (1) a shock breakout flash, (2) a stage of planar shock breakout cooling, and finally (3) shock cooling emission from the thermal energy released by the collision. The planar phase is unique to the unusual density profile of the He-detonated layer in comparison to the steep profile at a stellar edge as is usually considered for shock breakout. The shock cooling emission can be modified by recombination, and we explore these effects. All together, we expect an initial flash dominated by the planar phase of $\sim6\times10^{43}\,{\rm erg\,s^{-1}}$, which lasts $\sim5\,{\rm s}$ in the soft X-rays. This is followed by $\sim12-24\,{\rm hrs}$ of shock cooling at a luminosity of $3-10\times10^{40}\,{\rm erg\,s^{-1}}$ in the optical/UV. We discuss prospects for detection of this early DD emission with current and upcoming surveys.
\end{abstract}


\section{Introduction}
\label{sec:introduction}

Type Ia supernovae (SNe Ia) are crucial for our understanding of nucleosynthesis, galaxy evolution, and as cosmological probes. They are generally accepted as the thermonuclear destruction of CO white dwarfs \citep[WDs, see][]{Maoz2014}, but the details of their progenitors and the explosion mechanisms are still debated.

One promising scenario is the double detonation (DD). In a DD, a surface He layer first undergoes thermonuclear runaway. This then triggers a secondary explosion when the inward propagating shock from the initial explosion converges at some location within the CO core \citep{Livne1991,Woosley1994,Fink2007,Fink2010,Shen2014}. DDs were historically disfavored because early models focused on relatively thick He layers, which were shown to not match the majority of SNe~Ia \citep{Woosley1994,Hoeflich1996,Nugent1997}. Later work using thinner He shells (and a more careful treatment of the He-layer burning, which was less complete than previously thought) could produce spectroscopically normal SNe~Ia \citep{Kromer2010,Woosley2011,Townsley2012,Townsley2019} as well as naturally explains the primary parameter behind the Phillips relation as being the exploding CO WD mass \citep{Shen2021}.

DD models come in many varieties. The earliest models simply considered accretion from a degenerate or nondegenerate He-rich companion or the stable burning of accreted hydrogen which leads to a He-rich surface layer \citep[e.g.,][]{Taam1980,Nomoto1982,Woosley1986}. It was later shown that smaller amounts of He could still ignite dynamically when  transferred in a double degenerate system from the less massive WD to the surface of the more massive CO WD \citep{Guillochon2010}, known as the dynamically driven double degenerate double detonation (D$^6$) model \citep{Shen2018a}. In this case, the less massive WD could be He or CO since even the latter are born with sufficiently large surface He layers. The SN Ia explosion could result in unbinding the binary system, and runaway stars have been identified, potentially confirming this picture \citep{Shen2018,Werner2024}. In other cases the companion WD might not be ejected but instead also triggered by the first SN Ia, leading to a triple or quadruple detonation \citep{Papish15,Tanikawa19,Pakmor22,Boos2024}. There may be evidence for this via the bimodal nebular lines observed for some SNe~Ia \citep{Tucker2025}. Yet another way a DD could occur in double degenerate systems is if the impacting accretion stream reaches high enough temperatures and densities during mass transfer, because it may then directly ignite the He-rich surface layer of a CO WD \citep{Shen2024}.

Given the successes and continued interest in DD scenarios in recent years, it is natural to ask if there are predictions that might help distinguish it from other models. Motivated by this, we focus on the early emission expected from the collision of the fast outer material of the CO detonation with the previously detonated He layer. This leads to three main features, (1) a shock breakout, (2) a stage of planar shock breakout cooling, and (3) later shock cooling emission. We find that the shock breakout is likely too dim to be observed, but the other emission features are generally brighter than expected from the explosion of a bare WD because the collision naturally occurs at larger radii. 

In Section~\ref{sec:model}, we summarize the basic features of our model and the resulting three emission features. In Section~\ref{sec:prospects}, we discuss the prospects for observing these features with current and upcoming surveys. We conclude in Section~\ref{sec:conclusion} with a summary of our results and a discussion of future work on this topic.

\section{Collision Emission Features}
\label{sec:model}

\begin{figure}
\includegraphics[width=0.45\textwidth,trim=1.0cm 1.0cm 0.5cm 0.5cm]{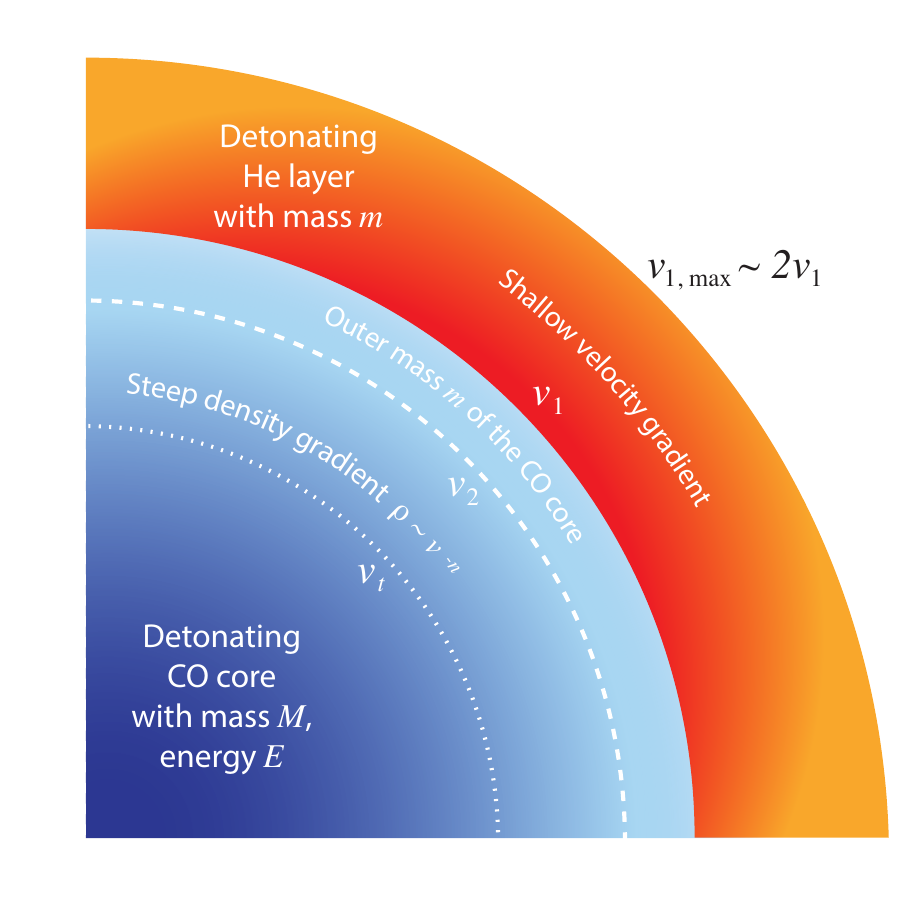}
\caption{Diagram of a quadrant of a DD, highlighting the key velocities considered in this model. The total system is broken into two regions, (1) the detonating He surface layer with mass $m$ (orange) and (2) the detonating CO WD with mass $M$ (blue). For velocities $v>v_t$ (dotted line), the core is imparted with a steep gradient $\rho\propto v^{-n}$ due to the detonation shock propagating into the edge of the star. In contrast, the prior He detonation moves roughly perpendicular in the layer and thus produces a shallower velocity gradient with $v_{1,\rm max}\sim2 v_1$, where $v_1\approx1.3\times10^{9}\,{\rm cm\,s^{-1}}$ is the characteristic velocity of the layer. During the collision, roughly only the outer $m$ of the CO core participates, which defines the velocity $v_2$ (dashed line).}
\label{fig:diagram}
\end{figure}

We consider the DD of a CO WD with mass $M$ with a small He surface layer with mass $m$. The key velocities considered in this process are summarized in Figure~\ref{fig:diagram} to help guide the reader, which we next describe below in some detail.

The He layer ignites first and expands with a characteristic velocity of $v_1\approx (2\epsilon_{\rm He})^{1/2} \approx1.3\times10^9\,{\rm cm\,s^{-1}}$, where $\epsilon_{\rm He}=5.8\times10^{17}\,{\rm erg\,g^{-1}}$ is the energy production from He burning. This estimate for $v_1$ assumes fairly complete burning of He. If the He-burning is less complete \citep[e.g.,][]{Boos2021}, $v_1$ can decrease somewhat but this is helped by $v_1$ depending on the square root of the energy production. This generates a shock that propagates into the CO core, eventually igniting a second detonation, which releases an energy $E$ and unbinds the WD at a time $\Delta t$ after the ignition of the He layer. The time delay $\Delta t$ is the sum of three consecutive processes. First is the time it takes the He shell detonation wave to engulf a significant fraction of the progenitor. This detonation wave moves at about $v_1$ and must travel halfway around the WD, setting this time at $\approx 1.5(M/M_\odot)^{-1/3}\,{\rm s}$. Second is the time it takes the compression wave to cross the WD, which is comparable to the sound crossing time of $\approx 1(M/M_\odot)^{-1}\,{\rm s}$. Finally, the time it takes the detonation wave to cross the exploding WD is $\approx 0.5(M/M_\odot)^{-1/3}\,{\rm s}$. We therefore expect $\Delta t$ to be a few seconds. This is consistent with more detailed numerical simulations that find a delay of $\Delta t\approx2-3\,{\rm s}$ \citep[e.g.,][and private communication with S.~J.~Boos]{Boos2021}, although with a dependence on $m$ that is not captured by these simple scalings.

The CO detonation generates a steep density profile in the outer layers of the CO WD of roughly \citep{Chevalier89}
\be
    \rho(v,t)
        = \frac{KM}{v_t^3t^3}
            \lp \frac{v}{v_t} \rp^{-n},
        \label{eq:rho}
\ee
where mass conservation sets
\be
    K = \frac{(n-3)(3-\delta)}{4\pi(n-\delta)},
\ee
with typical values of $n\approx10$ and $\delta\approx1.1$ (which gives $K=0.119$). The transition velocity (denoted by a dotted line in Figure~\ref{fig:diagram})
\be
    v_t
    &=&
    \lb \frac{(n-5)(5-\delta)}{(n-3)(\delta-3)}\rb^{1/2}
        \lp \frac{2E}{M} \rp^{1/2}
    \nonumber
    \\
    &=&1.2\times10^9
        M_1^{-1/2}E_{51}^{1/2}\,{\rm cm\,s^{-1}},
\ee
where $M_1=M/M_\odot$ and $E_{51}=E/10^{51}\,{\rm erg}$. Although a sub-Chandrasakhar SN Ia is exploded by a detonation wave, below a density of $\approx2\times10^6\,{\rm g\,cm^{-3}}$ the detonation will transition to a shock \citep[as discussed in more detail by][]{Piro2010}. This shock accelerates near the stellar edge in a manner that is similar to the core-collapse case \citep[e.g.,][]{Sakuria1960} which justifies the density profile used in Equation~
(\ref{eq:rho}).

During the collision, roughly the outer mass $m$ of the CO WD will interact with the He detonation shell. The velocity at this mass roughly sets the velocity $v_2$ of the collision (denoted as a dashed line in Figure~\ref{fig:diagram}),
\be
    m(v_2)
        &=& \int_{v_2}^\infty 4\pi (vt)^2\rho(v,t)d(vt)
        \nonumber
        \\
        &=& \frac{4\pi KM}{n-3}
            \lp \frac{v_2}{v_t} \rp^{-n+3}. 
\ee
Solving for $v_2$ and adding an extra factor of $2$ due to shock acceleration from pressure gradients \citep{Matzner99} results in
\be
    v_2 &=& 2v_t
        \lp \frac{n-3}{4\pi K}\rp^{1/(3-n)}
        \lp \frac{m}{M} \rp^{1/(3-n)}
    \nonumber
    \\
    &=& 3.7\times10^9
        m_{-2}^{-0.14}M_1^{-0.36}E_{51}^{1/2}\,
        {\rm cm\,s^{-1}},
    \label{eq:v2}
\ee
where $m_{-2}=m/10^{-2}\,M_\odot$.

\subsection{Shock Breakout}

The radius of the detonated He layer at a time $t$ after the CO core detonates is $v_1(t+\Delta t)$ while the radius of the outer CO ejecta is $v_2t$. Thus they will collide at a time $t_{\rm col} = v_1\Delta t/(v_2-v_1)$ and at a radius
\be
    r_{\rm col} = \frac{v_2v_1\Delta t}{v_2-v_1} .
\ee
The collision causes a shock with a velocity of roughly $v_2$ to travel outward through the detonated He layer. The propagation of shocks through an expanding medium was explored by \citet{Govreen-Sega2024}. For our treatment, here we approximate these more complicated effects, focusing on the observable signatures.

The detonated He shell will have a velocity profile varying from roughly $v_1$ to a maximum velocity of $v_{1,\rm max}$. For shock breakout to occur requires $v_2>v_{1,\rm max}$. Unlike the detonation shock that unbinds the WD, where the shock propagates in a decreasing density profile, the earlier shock that crosses He layer moves in a direction that is roughly perpendicular to the layer density gradient. Therefore, the entire surface layer is accelerated by the shock roughly to a similar velocity, where the velocity gradient in the layer is set mostly by the rarefaction wave that crosses the shocked layer. Such a wave typically results in a density/velocity gradient without a tail to low densities and high velocities. We therefore consider $v_{1,\rm max}\sim2v_1$, but we keep in mind that in principle $v_{1,\rm max}$ could be higher, and shock breakout might not even happen. 

Nevertheless, if $v_2>v_{1,\rm max}$, then the shock will overtake the fastest moving material at a radius
\be
    r_{\rm bo} \approx \frac{v_2v_{1,\rm max}\Delta t}{v_2-v_{1,\rm max}}
    = 1.5 \times 10^{10} \eta_{0.4}^{-1}  \lp\frac{v_{\rm 1,max}\Delta t}{6\times10^9\,{\rm cm}}\rp {\rm\,cm},
    \nonumber
    \\
\ee
where we define $\eta \equiv 1-v_{1,\rm max}/v_2$ as a parameter that reflects the uncertainty in the exact value of $v_{1,\rm max}$ and  $\eta_{0.4}=\eta/0.4$.
Since the breakout occurs at an optical depth $\tau\approx c/(v_2-v _{1,\rm max})$, it has an associated mass
\be
    m_{\rm bo}
    &\approx& \frac{4\pi r_{\rm bo}^2}{\kappa}
        \frac{c}{v_2-v _{1,\rm max}}
    \nonumber
        \\
    &\approx& 3 \times 10^{-10}
        \kappa_{0.1}^{-1}
        m_{-2}^{0.14}M_1^{0.36}E_{51}^{-1/2}
        \nonumber
        \\
        &&\times
        \lp\frac{v_{\rm 1,max}\Delta t}{6\times10^9\,{\rm cm}}\rp^2 \eta_{0.4}^{-3}\,M_\odot,
\ee
where $\kappa$ is the opacity and $\kappa_{0.1}=\kappa/0.1\,{\rm cm^2\,g^{-1}}$. We then estimate
\be
    E_{\rm bo} &\approx& \frac{m_{\rm bo}}{2}(v_2-v _{1,\rm max})^2
    \nonumber
        \\
    &\approx& 6\times10^{41}
        \kappa_{0.1}
        m_{-2}^{-0.14}M_1^{-0.36}E_{51}^{1/2}
        \nonumber
        \\
        &&\times
        \lp\frac{v_{\rm 1,max}\Delta t}{6\times10^9\,{\rm cm}}\rp^2 \eta_{0.4}^{-1}\,{\rm\,erg},
\ee
for the energy of the breakout pulse.

If the breakout is spherical, this energy is observed over a duration of $r_{\rm bo}/c$, which is a fraction of a second. However, the expanding He shell is not expected to be spherical due to the transverse detonation it experiences, and thus the observed energy will be spread over a longer time. A rough upper limit of the flash duration is $r_{\rm bo}/v_{1,\rm max}$, which is of order of $10\,{\rm s}$ (assuming that $v_2>v_{1,\rm max}$ and there is a breakout). The breakout emission is in thermal equilibrium unless $v_2-v_{1,\rm max}>0.1c$ \citep[]
[note that this limit is higher than in the case of a breakout from a stellar edge, due to the much higher densities for the case here]{Katz2010}. We therefore estimate an observed breakout temperature of
\be
    T_{\rm bo} &\approx& \lp\frac{E_{\rm bo} c}{4\pi\sigma_{\rm SB} r_{\rm bo}^3}\rp^{1/4}
     \nonumber
        \\
    &\approx& 2 \times 10^6 \kappa_{0.1}^{-1/4}
        m_{-2}^{-0.035}M_1^{-0.09}E_{51}^{1/8}
        \nonumber
        \\
        &&\times
        \lp\frac{v_{\rm 1,max}\Delta t}{6\times10^9\,{\rm cm}}\rp^{1/4} \eta_{0.4}^{1/2}\, {\rm\,K},
\ee
where $\sigma_{\rm SB}$ is the Stefan-Boltzmann constant.

%

\subsection{Planar Phase}

In most circumstances, the shock breakout may be too dim and short-lived to observe. Nevertheless, a key difference that we have been emphasizing between shock breakout here and typical shock breakout from many supernovae is that for a DD there is no steep velocity gradient at the edge of the expanding He layer. In a typical breakout from a stellar surface, the steep density gradient implies that the breakout timescale (set by the diffusion through the breakout layer) is comparable to the dynamical timescale of the breakout layer (set by the time the breakout layer doubles its width). As a result, during the entire planar phase (the time during which the breakout layer doubles its radius), the radiation can diffuse out only from the breakout layer, and the diffusion wave is roughly at a constant Lagrangian coordinate. In contrast, the shallow density velocity profile of the detonated He layer, implies that the dynamical time of the breakout layer is much longer than the diffusion time during the breakout. This means that during the early planar phase of expansion (timescales $\lesssim r_{\rm bo}/v_2$) of a DD, the diffusion wave can move inward (in a Lagrangian sense) and can access additional shock-heated material (this is similar to the a breakout process that may take place in the ejecta of binary neutron star merges, see \citealp{Nakar2020} for details). We consider the effects of this in more detail next.

If we donate $f$ as the fraction of shell material that has a velocity near $\sim v_{1,\rm max}$, then the density of that material at the time of breakout is
\be
    \rho_{\rm bo} &\approx& \frac{fm}{4\pi r_{\rm bo}^3}.
\ee
If a width of material $d_{\rm pl}$ cools during the planar phase, its diffusion time is roughly $\kappa\rho_{\rm bo}d_{\rm pl}^2/c$. Setting this to the planar timescale of $r_{\rm bo}/v_2$, we find
\be
    d_{\rm pl} \approx
    \lp \frac{4\pi r_{\rm bo}^4}{\kappa fm}
        \frac{c}{v_2}\rp^{1/2}.
\ee
Thus the mass of material that can cool during the planar phase is
\be
    m_{\rm pl} \approx
    4\pi r_{\rm bo}^2 d_{\rm pl} \rho_{\rm bo}
    \approx \lp \frac{4\pi r_{\rm bo}^2 fm}{\kappa}\frac{c}{v_2}\rp^{1/2}.
\ee
Interestingly, this can alternatively be expressed as the geometric mean of the mass of material with velocity near $v_{1,\rm max}$ and the breakout mass, $m_{\rm pl}\approx (f m m_{\rm bo})^{1/2}$. The associated energy is
\be
    E_{\rm pl}
    \approx \frac{m_{\rm pl}}{2}(v_2-v _{1,\rm max})^2 .
\ee
Dividing this energy by the planar timescale $r_{\rm bo}/v_2$ gives a planar luminosity of
\be
    L_{\rm pl}
    &\approx& \lp\frac{\pi fm}{\kappa}\frac{c}{v_2}\rp^{1/2}
    \lp v_2-v_{1,\rm max}\rp^2v_2
    \nonumber
    \\
    &\approx& 6\times10^{43}
    \kappa_{0.1}^{-1/2}f_{-2}^{1/2}m_{-2}^{0.65}
    M_1^{-0.9}E_{51}^{5/4}\eta_{0.4}^2\,{\rm erg\,s^{-1}},
    \nonumber
    \\
\ee
where $f_{-2}=f/10^{-2}$, which will last for a timescale of
\be
    t_{\rm pl} &\approx& \frac{r_{\rm bo}}{v_2}
    \approx \frac{v_{1,\rm max}\Delta t}{v_2-v_{1,\rm max}}
    \nonumber
    \\
    &\approx& 4\,
    m_{-2}^{0.14}M_1^{0.36}E_{51}^{-1/2}
    \lp\frac{v_{\rm 1,max}\Delta t}{6\times10^9\,{\rm cm}}\rp\eta_{0.4}^{-1}\,{\rm s}.
\ee
The exact temperature of this emission depends on the degree of coupling. A lower limit can be estimated from $T_{\rm pl}\approx (L_{\rm pl}/4\pi\sigma_{\rm SB}r_{\rm bo}^2)^{1/4}$, which results in $T_{\rm pl}\approx4\times10^6\,{\rm K}$. We conclude that if a shock breakout occurs, then the planar phase will be the most readily detectable signal.

\subsection{Shock Cooling Emission}
\label{sec:sce}

The final stage of emission is due to the cooling of shock-heated material as it expands and radiates. This will occur even in cases where there is no shock breakout or planar phase because $v_2<v_{1,\rm max}$. The collision of the fast-moving material with velocity $v_2$ with the slower-moving layer with velocity $v_1$ releases thermal energy
\be
    E_{\rm col} \approx \frac{m}{4}(v_2-v_1)^2,
\ee
where we assume a prefactor of $1/4$ but the exact value depends on the details of the collision (this factor comes from considering an inelastic collision of two slabs of equal mass $m$). In the limit $v_2\gg v_1$\footnote{Throughout this subsection we use the $v_2\gg v_1$ limit when presenting numerical estimates.},
\be
    E_{\rm col} \approx 7\times10^{49}m_{-2}^{0.72}M_1^{-0.72}E_{51}\,{\rm erg}.
    \label{eq:deltae}
\ee
Note that Equations (\ref{eq:v2}) and (\ref{eq:deltae}) are similar to what was found by \citet{Nakar2014} but in the different context of the explosion of yellow supergiants with low-mass extended envelopes.

Unlike the shock breakout and planar phase, this energy is not seen directly by an observer. Instead, an observer sees the shock-heated shell once it has expanded sufficiently that its optical depth satisfies $\tau \approx c/v_f$ where $v_f\approx (v_1+v_2)/2$ is the final velocity of the shell after the collision. This results in an observed shock cooling radius of
\be
    r_{\rm sc} &\approx & \lp \frac{\kappa m v_f}{4\pi c}\rp^{1/2}.
    \nonumber
    \\
    &\approx& 9 \times10^{13}
        \kappa_{\rm 0.1}^{1/2}m_{-2}^{0.43}M_1^{-0.18}E_{51}^{1/4}\,{\rm cm}.
\ee
The timescale associated with reaching this radius is
\be
    t_{\rm sc} &\approx& r_{\rm sc}/v_f
    \nonumber
    \\
    &\approx& 15\,\kappa_{0.1}^{1/2}m_{-2}^{0.57}M_1^{0.18}E_{51}^{-1/4}\,{\rm hrs}.
\ee
Adiabatic expansion decreases the the energy by a factor of $r_{\rm col}/r_{\rm sc}$ during the time it takes photons to escape the shock heated region. Thus, the observed luminosity is
\be
    L_{\rm sc} &=& \frac{E_{\rm col}}{t_{\rm sc}}
                \lp \frac{r_{\rm col}}{r_{\rm sc}}\rp
    \nonumber
    \\
    &\approx& 4\times10^{40}
        \kappa_{0.1}^{-1}m_{-2}^{-0.28}M_1^{-0.72}E_{51}
        \nonumber
        \\
        &&\times\lp\frac{v_1\Delta t}{3\times10^9\,{\rm cm}}\rp
        {\rm erg\,s^{-1}},
\ee
with
\be
    T_{\rm sc} &=&
    \lp \frac{L_{\rm sc}}{4\pi \sigma_{\rm SB}r_{\rm sc}^2}\rp^{1/4}
    \nonumber
    \\
    &\approx& 9,000\,\kappa_{0.1}^{-1/2}m_{-2}^{-0.29}
        M_1^{-0.09}E_{51}^{1/8}
        \nonumber
        \\
        &&\times\lp\frac{v_1\Delta t}{3\times10^9\,{\rm cm}}\rp^{1/4}
        {\rm K},
\ee
as the characteristic shock cooling temperature.

\begin{figure}
\includegraphics[width=0.45\textwidth,trim=1.0cm 5.0cm 1.7cm 2.8cm]{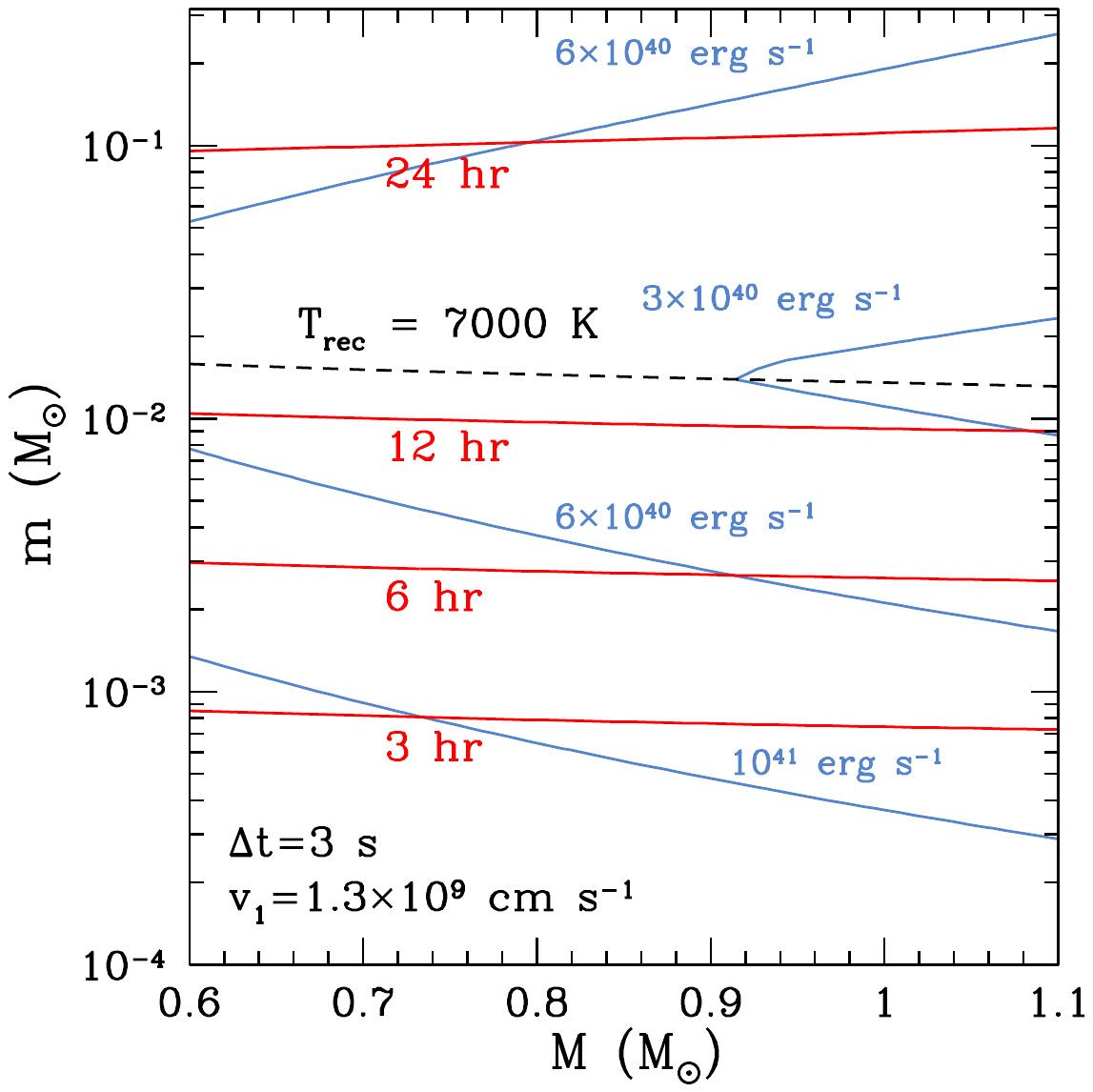}
\caption{Contours of constant luminosity (blue) and timescale (red) expected for shock cooling emission from a DD as a function of the WD mass $M$ and the He surface layer mass $m$. The initial He-detonation velocity is set to $v_1=1.3\times10^9\,{\rm cm\,s^{-1}}$ with a delay of $\Delta t=3\,{\rm s}$ from the initiation of the He detonation to the unbinding of the core. We fix the explosion energy of the core detonation to $E=10^{51}\,{\rm erg}$ and use an opacity of $\kappa=0.1\,{\rm cm^2\,g^{-1}}$ for the cooling emission. Recombination occurs above the dashed black line for $T_{\rm rec}=7,000,{\rm K}$. Unlike the numerical scalings summarized in Section~\ref{sec:sce}, we do not approximate $v_2\gg v_1$ for this plot.}
\label{fig:luminosity_time}
\end{figure}

This temperature justifies our choice of opacity for most of the parameter range. Although at sufficiently large He shell masses $m\gtrsim0.02\,M_\odot$, then $T_{\rm obs}\lesssim7,000\,{\rm K}$ and the ashes from the He detonation will start recombining. This causes the outer layers of the mass $m$ to become effectively transparent to the thermal emission from the shock heated material \citep[see examples of the opacity's dependence on temperature in][]{Piro2014b}. We treat this by replacing the shock cooling luminosity from above with
\be
    L_{\rm rec} = \frac{E_{\rm col}}{r_{\rm rec}/v_f}
    \lp \frac{r_{\rm col}}{r_{\rm rec}}\rp,
    \label{eq:lrec}
\ee
where $r_{\rm rec}$ is the radius where recombination occurs. This radius is set by
\be
    T_{\rm rec} = \lp \frac{L_{\rm rec}}{4\pi \sigma_{\rm SB}r_{\rm rec}^2}\rp^{1/4}.
    \label{eq:temprec}
\ee
This approximation is reasonable since once the recombination wave reaches the photosphere radius, $r_{\rm rec}$, the luminosity satisfies $\approx 4\pi\sigma_{\rm SB}r_{\rm rec}^2 T_{\rm rec}^4$ until most of the internal energy in the shell is radiated (similar to the plateau phase of a Type IIP supernova). Combining Equations (\ref{eq:lrec}) and (\ref{eq:temprec}), we  solve for the recombination radius for a given recombination temperature $T_{\rm rec}$,
\be
    r_{\rm rec} &=&
    \lp\frac{E_{\rm col} v_f r_{\rm col}}{4\pi\sigma_{\rm SB}T_{\rm rec}^4} \rp^{1/4}
    \nonumber
    \\
    &\approx& 1\times10^{14}
        m_{-2}^{0.14}M_1^{-0.27} E_{51}^{3/8}
        \nonumber
        \\
        &&\times\lp\frac{v_1\Delta t}{3\times10^9\,{\rm cm}}\rp^{1/4} \lp\frac{T_{\rm rec}}{7,000\,{\rm K}}\rp^{-1}{\rm cm}.
\ee
An observer will now see the shock cooling on the timescale for the recombination to occur is
\be
    t_{\rm rec} &=& r_{\rm rec}/v_f
    \nonumber
    \\
    &\approx& 18\,
    m_{-2}^{0.28}M_1^{0.090} E_{51}^{-1/8}
        \nonumber
        \\
        &&\times\lp\frac{v_1\Delta t}{3\times10^9\,{\rm cm}}\rp^{1/4} \lp\frac{T_{\rm rec}}{7,000\,{\rm K}}\rp^{-1}{\rm hr},
\ee
with an observed luminosity
\be
    L_{\rm rec} &\approx& 2\times10^{40}
    m_{-2}^{0.28}M_1^{-0.54} E_{51}^{3/4}
        \nonumber
        \\
        &&\times\lp\frac{v_1\Delta t}{3\times10^9\,{\rm cm}}\rp^{1/2} \lp\frac{T_{\rm rec}}{7,000\,{\rm K}}\rp^{2}
        {\rm erg\,s^{-1}},
\ee
Basically, in cases where $T_{\rm sc}\lesssim T_{\rm rec}$, recombination causes the observed radius to be smaller. This in turn allows the shock cooling to be observed earlier when less adiabatic cooling has taken place so that the luminosity is higher.

The results of our estimates for the shock cooling expected at early times from a DD are summarized in Figure~\ref{fig:luminosity_time}. This includes lines of constant luminosity and timescale (in blue and red, respectively). The black dashed line denotes where the temperature falls below the recombination temperature (chosen to be $T_{\rm rec}=7,000\,{\rm K}$ here). This demonstrates that above the recombination line the shock cooling starts becoming brighter with increasing $m$, while the timescale starts to depend less strongly on $m$ so that it is typically $\sim1\,{\rm day}$.

\section{Prospects for Detection}
\label{sec:prospects}

The previous section shows that shock breakout and shock cooling from a DD may produce an early observable signal. Whether or not the shock breakout occurs depends on a more detailed treatment that resolves the velocity profile of the detonated He shell and how the shock driven by the secondary CO detonation propagates through this already moving material. Our work demonstrates, though, that if the shock breakout occurs, it would be most readily detectable from the planar phase emission. The two most promising instruments for detecting this would be the {\it Neil Gehrels Swift Observatory} XRT (the burst would likely be too soft to be observed with the wider field BAT) and the {\it Einstein Probe} WXT. In either case, the signal could be detected out to $\sim20\,{\rm Mpc}$ where there are typically a $\sim\,$few SNe~Ia per year. Detection with {\it Swift} would almost have to be serendipitous, but the wider field of view of the {\it Einstein Probe} may allow a couple of DD SNe Ia to be detected for a five year mission lifetime.

The subsequent shock cooling signal is dimmer but longer lasting, which may make it easier to observe. Typical absolute magnitudes would be in the range of $-$12.5 up to $-$13.5, which could last up to $\sim1\,{\rm day}$ for the full rise and fall. Indeed, even fairly normal SNe~Ia show significant diversity when the focus is on the early evolution. The first collections of early SN~Ia photometry exhibited a dichotomy of red- and blue-early emission \citep{Stritzinger2018}, and more recently it has been suggested that there may be a range of early colors or even more groups \citep{Burke2022,Hoogendam2024,Ni2024}. Even with this growing sample of early events, there is still very few detected within $\sim1\,{\rm day}$, and many of the events where there is an indication of an excess early flux, the signal is too large to be explained by DD shock cooling \citep[e.g.,][]{Dimitriadis2019,Xi2024}.

This situation should drastically improve with the launch of {\it Ultrasat} \citep{BenAmi2022,Shvartzvald2024}. Its $204\,{\rm deg^2}$ field of view coupled with its focus on the near-ultraviolet are ideally suited for studying these early phases. The $10\,{\rm min}$ exposures of the low-cadence survey should be sensitive to these features out to $\sim50\,{\rm Mpc}$, which should catch a $\sim$ few events per year given the wide field of view. The largest challenge will be distinguishing DD shock cooling from the many other early emission processes that have been predicted for SNe Ia. This includes, just to name a few, shock cooling from a bare WD surface itself \citep{Piro2010}, collision with a non-degenerate companion \citep{Kasen2010,Maeda2014,Kutsuna2015,Liu2015}, shallow mixing of radioactive nickel \citep{Piro2014a,Piro2016,Noebauer2017,Maeda2018,Polin2019,Magee2020}, interaction with circumstellar material \citep{Piro2016,Kromer2016,Levanon2019,Moriya2023}. In some cases, Doppler shifts of certain features may event generate a seeming excess in a given band \citep{Ashall2022}. Ultimately, connecting these early features with other properties of the SN \citep[e.g., its SN~Ia classification, luminosity, host galaxy properties, features seen at other phases, see][]{Hoogendam2024} will be key for putting together a more complete picture to identify which events are likely DDs. Having a larger sample of early SN~Ia detections will be a critical step in this direction.

\section{Conclusion and Discussion}
\label{sec:conclusion}

We explored the early emission signatures expected from a DD due to the collision of sub-Chandrasekhar CO detonation with the previously detonated surface He shell. We identify three possible features. The first is shock breakout, which lasts $\lesssim10\,{\rm s}$ because the breakout is likely not spherical due to the transverse detonation that first explodes the He layer. It will likely be fairly dim at $\lesssim\times10^{41}\,{\rm erg\,s^{-1}}$. Next is shock breakout planar cooling, which is unique due to the He shell having a density profile with a shallower gradient in comparison to a typical steep stellar profile. This gives a larger luminosity of $\sim6\times10^{43}\,{\rm erg\,s^{-1}}$ and lasts $\sim4\,{\rm s}$, which is expected to peak in soft X-rays. Finally there is shock cooling emission that lasts for $\sim12-24\,{\rm hrs}$ at a luminosity of $3-10\times10^{40}\,{\rm erg\,s^{-1}}$ (as summarized in Figure~\ref{fig:luminosity_time}). Our work also includes a novel treatment of recombination during the shock cooling phase, which is due to the intermediate mass elements that dominate the ashes of the He-detonated layer. The collision can also re-trigger burning in the low-velocities regions of the helium ashed, which may produce interesting chemical signatures seen in the shock cooling phase, especially for lines of sight closest to the helium ignition point \citep[e.g.,][]{Boos2021}.

Future, more detailed work on the propagation of the CO detonation shock through the He shell is needed to better understand when shock breakout occurs and the details of its emission properties. Although there is a growing collection of early-detected SNe~Ia, many are still not early enough and/or at sufficient depth to test the occurrence of this emission. There are also predictions for many other emission features that could mimic or even be much brighter than what we predict here in the optical/UV. Nevertheless, {\it Ultrasat} is poised to provide a data set that will finally start to help answer some of these questions in the coming years. In X-ray, there is no prediction of another process that results in a signal that is similar to the planar emission we predict here. Therefore, detection of such a signal, e.g., by the Einstein Probe \citep{Yuan15}, from SNe Ia will provide support to the DD model.

\acknowledgments
We thank Samuel Boos, Ken Shen, and Yossef Zenati for providing feedback on a previous draft. EN was partially supported by 
a consolidator ERC grant 818899 (JetNS) and an ISF grant (1995/21).

\bibliographystyle{aasjournal}

\begin{thebibliography}{}
\expandafter\ifx\csname natexlab\endcsname\relax\def\natexlab#1{#1}\fi
\providecommand{\url}[1]{\href{#1}{#1}}
\providecommand{\dodoi}[1]{doi:~\href{http://doi.org/#1}{\nolinkurl{#1}}}
\providecommand{\doeprint}[1]{\href{http://ascl.net/#1}{\nolinkurl{http://ascl.net/#1}}}
\providecommand{\doarXiv}[1]{\href{https://arxiv.org/abs/#1}{\nolinkurl{https://arxiv.org/abs/#1}}}

\bibitem[{{Ashall} {et~al.}(2022){Ashall}, {Lu}, {Shappee}, {Burns}, {Hsiao}, {Kumar}, {Morrell}, {Phillips}, {Shahbandeh}, {Baron}, {Boutsia}, {Brown}, {DerKacy}, {Galbany}, {Hoeflich}, {Krisciunas}, {Mazzali}, {Piro}, {Stritzinger}, \& {Suntzeff}}]{Ashall2022}
{Ashall}, C., {Lu}, J., {Shappee}, B.~J., {et~al.} 2022, \apjl, 932, L2, \dodoi{10.3847/2041-8213/ac7235}

\bibitem[{{Ben-Ami} {et~al.}(2022){Ben-Ami}, {Shvartzvald}, {Waxman}, {Netzer}, {Yaniv}, {Algranatti}, {Gal-Yam}, {Lapid}, {Ofek}, {Topaz}, {Arcavi}, {Asif}, {Azaria}, {Bahalul}, {Barschke}, {Bastian-Querner}, {Berge}, {Berlea}, {Buehler}, {Dittmar}, {Gelman}, {Giavitto}, {Guttman}, {Haces Crespo}, {Heilbrunn}, {Kachergincky}, {Kaipachery}, {Kowalski}, {Kulkarni}, {Kumar}, {K{\"u}sters}, {Liran}, {Miron-Salomon}, {Mor}, {Nir}, {Nitzan}, {Philipp}, {Porelli}, {Sagiv}, {Schliwinski}, {Sprecher}, {De Simone}, {Stern}, {Stone}, {Trakhtenbrot}, {Vasilev}, {Watson}, \& {Zappon}}]{BenAmi2022}
{Ben-Ami}, S., {Shvartzvald}, Y., {Waxman}, E., {et~al.} 2022, in Society of Photo-Optical Instrumentation Engineers (SPIE) Conference Series, Vol. 12181, Space Telescopes and Instrumentation 2022: Ultraviolet to Gamma Ray, ed. J.-W.~A. {den Herder}, S.~{Nikzad}, \& K.~{Nakazawa}, 1218105, \dodoi{10.1117/12.2629850}

\bibitem[{{Boos} {et~al.}(2024){Boos}, {Townsley}, \& {Shen}}]{Boos2024}
{Boos}, S.~J., {Townsley}, D.~M., \& {Shen}, K.~J. 2024, \apj, 972, 200, \dodoi{10.3847/1538-4357/ad5da2}

\bibitem[{{Boos} {et~al.}(2021){Boos}, {Townsley}, {Shen}, {Caldwell}, \& {Miles}}]{Boos2021}
{Boos}, S.~J., {Townsley}, D.~M., {Shen}, K.~J., {Caldwell}, S., \& {Miles}, B.~J. 2021, \apj, 919, 126, \dodoi{10.3847/1538-4357/ac07a2}

\bibitem[{{Burke} {et~al.}(2022){Burke}, {Howell}, {Sand}, {Amaro}, {Brown}, {Andrews}, {Bostroem}, {Dong}, {Haislip}, {Hiramatsu}, {Hosseinzadeh}, {Kouprianov}, {Lundquist}, {McCully}, {Pellegrino}, {Reichart}, {Tartaglia}, {Valenti}, \& {Yang}}]{Burke2022}
{Burke}, J., {Howell}, D.~A., {Sand}, D.~J., {et~al.} 2022, arXiv e-prints, arXiv:2207.07681, \dodoi{10.48550/arXiv.2207.07681}

\bibitem[{{Chevalier} \& {Soker}(1989)}]{Chevalier89}
{Chevalier}, R.~A., \& {Soker}, N. 1989, \apj, 341, 867, \dodoi{10.1086/167545}

\bibitem[{{Dimitriadis} {et~al.}(2019){Dimitriadis}, {Foley}, {Rest}, {Kasen}, {Piro}, {Polin}, {Jones}, {Villar}, {Narayan}, {Coulter}, {Kilpatrick}, {Pan}, {Rojas-Bravo}, {Fox}, {Jha}, {Nugent}, {Riess}, {Scolnic}, {Drout}, {K2 Mission Team}, {Barentsen}, {Dotson}, {Gully-Santiago}, {Hedges}, {Cody}, {Barclay}, {Howell}, {KEGS}, {Garnavich}, {Tucker}, {Shaya}, {Mushotzky}, {Olling}, {Margheim}, {Zenteno}, {Kepler spacecraft Team}, {Coughlin}, {Van Cleve}, {Cardoso}, {Larson}, {McCalmont-Everton}, {Peterson}, {Ross}, {Reedy}, {Osborne}, {McGinn}, {Kohnert}, {Migliorini}, {Wheaton}, {Spencer}, {Labonde}, {Castillo}, {Beerman}, {Steward}, {Hanley}, {Larsen}, {Gangopadhyay}, {Kloetzel}, {Weschler}, {Nystrom}, {Moffatt}, {Redick}, {Griest}, {Packard}, {Muszynski}, {Kampmeier}, {Bjella}, {Flynn}, {Elsaesser}, {Pan-STARRS}, {Chambers}, {Flewelling}, {Huber}, {Magnier}, {Waters}, {Schultz}, {Bulger}, {Lowe}, {Willman}, {Smartt}, {Smith}, {DECam}, {Points}, {Strampelli}, {ASAS-SN}, {Brimacombe}, {Chen}, {Mu{\~n}oz},
  {Mutel}, {Shields}, {Vallely}, {Villanueva}, {PTSS/TNTS}, {Li}, {Wang}, {Zhang}, {Lin}, {Mo}, {Zhao}, {Sai}, {Zhang}, {Zhang}, {Zhang}, {Wang}, {Zhang}, {Baron}, {DerKacy}, {Li}, {Chen}, {Xiang}, {Rui}, {Wang}, {Huang}, {Li}, {Cumbres Observatory}, {Hosseinzadeh}, {Howell}, {Arcavi}, {Hiramatsu}, {Burke}, {Valenti}, {ATLAS}, {Tonry}, {Denneau}, {Heinze}, {Weiland}, {Stalder}, {Konkoly}, {Vink{\'o}}, {S{\'a}rneczky}, {P{\'a}l}, {B{\'o}di}, {Bogn{\'a}r}, {Cs{\'a}k}, {Cseh}, {Cs{\"o}rnyei}, {Hanyecz}, {Ign{\'a}cz}, {Kalup}, {K{\"o}nyves-T{\'o}th}, {Kriskovics}, {Ordasi}, {Rajmon}, {S{\'o}dor}, {Szab{\'o}}, {Szak{\'a}ts}, {Zsidi}, {ePESSTO}, {Williams}, {Nordin}, {Cartier}, {Frohmaier}, {Galbany}, {Guti{\'e}rrez}, {Hook}, {Inserra}, {Smith}, {Arizona}, {Sand}, {Andrews}, {Smith}, \& {Bilinski}}]{Dimitriadis2019}
{Dimitriadis}, G., {Foley}, R.~J., {Rest}, A., {et~al.} 2019, \apjl, 870, L1, \dodoi{10.3847/2041-8213/aaedb0}

\bibitem[{{Fink} {et~al.}(2007){Fink}, {Hillebrandt}, \& {R{\"o}pke}}]{Fink2007}
{Fink}, M., {Hillebrandt}, W., \& {R{\"o}pke}, F.~K. 2007, \aap, 476, 1133, \dodoi{10.1051/0004-6361:20078438}

\bibitem[{{Fink} {et~al.}(2010){Fink}, {R{\"o}pke}, {Hillebrandt}, {Seitenzahl}, {Sim}, \& {Kromer}}]{Fink2010}
{Fink}, M., {R{\"o}pke}, F.~K., {Hillebrandt}, W., {et~al.} 2010, \aap, 514, A53, \dodoi{10.1051/0004-6361/200913892}

\bibitem[{{Govreen-Segal} {et~al.}(2024){Govreen-Segal}, {Youngerman}, {Palit}, {Nakar}, {Levinson}, \& {Bromberg}}]{Govreen-Sega2024}
{Govreen-Segal}, T., {Youngerman}, N., {Palit}, I., {et~al.} 2024, \mnras, 528, 313, \dodoi{10.1093/mnras/stad4000}

\bibitem[{{Guillochon} {et~al.}(2010){Guillochon}, {Dan}, {Ramirez-Ruiz}, \& {Rosswog}}]{Guillochon2010}
{Guillochon}, J., {Dan}, M., {Ramirez-Ruiz}, E., \& {Rosswog}, S. 2010, \apjl, 709, L64, \dodoi{10.1088/2041-8205/709/1/L64}

\bibitem[{{Hoeflich} \& {Khokhlov}(1996)}]{Hoeflich1996}
{Hoeflich}, P., \& {Khokhlov}, A. 1996, \apj, 457, 500, \dodoi{10.1086/176748}

\bibitem[{{Hoogendam} {et~al.}(2024){Hoogendam}, {Shappee}, {Brown}, {Tucker}, {Ashall}, \& {Piro}}]{Hoogendam2024}
{Hoogendam}, W.~B., {Shappee}, B.~J., {Brown}, P.~J., {et~al.} 2024, \apj, 966, 139, \dodoi{10.3847/1538-4357/ad33ba}

\bibitem[{{Kasen}(2010)}]{Kasen2010}
{Kasen}, D. 2010, \apj, 708, 1025, \dodoi{10.1088/0004-637X/708/2/1025}

\bibitem[{{Katz} {et~al.}(2010){Katz}, {Budnik}, \& {Waxman}}]{Katz2010}
{Katz}, B., {Budnik}, R., \& {Waxman}, E. 2010, \apj, 716, 781, \dodoi{10.1088/0004-637X/716/1/781}

\bibitem[{{Kromer} {et~al.}(2010){Kromer}, {Sim}, {Fink}, {R{\"o}pke}, {Seitenzahl}, \& {Hillebrandt}}]{Kromer2010}
{Kromer}, M., {Sim}, S.~A., {Fink}, M., {et~al.} 2010, \apj, 719, 1067, \dodoi{10.1088/0004-637X/719/2/1067}

\bibitem[{{Kromer} {et~al.}(2016){Kromer}, {Fremling}, {Pakmor}, {Taubenberger}, {Amanullah}, {Cenko}, {Fransson}, {Goobar}, {Leloudas}, {Taddia}, {R{\"o}pke}, {Seitenzahl}, {Sim}, \& {Sollerman}}]{Kromer2016}
{Kromer}, M., {Fremling}, C., {Pakmor}, R., {et~al.} 2016, \mnras, 459, 4428, \dodoi{10.1093/mnras/stw962}

\bibitem[{{Kutsuna} \& {Shigeyama}(2015)}]{Kutsuna2015}
{Kutsuna}, M., \& {Shigeyama}, T. 2015, \pasj, 67, 54, \dodoi{10.1093/pasj/psv028}

\bibitem[{{Levanon} \& {Soker}(2019)}]{Levanon2019}
{Levanon}, N., \& {Soker}, N. 2019, \apjl, 872, L7, \dodoi{10.3847/2041-8213/ab0285}

\bibitem[{{Liu} {et~al.}(2015){Liu}, {Moriya}, \& {Stancliffe}}]{Liu2015}
{Liu}, Z.-W., {Moriya}, T.~J., \& {Stancliffe}, R.~J. 2015, \mnras, 454, 1192, \dodoi{10.1093/mnras/stv2076}

\bibitem[{{Livne} \& {Glasner}(1991)}]{Livne1991}
{Livne}, E., \& {Glasner}, A.~S. 1991, \apj, 370, 272, \dodoi{10.1086/169813}

\bibitem[{{Maeda} {et~al.}(2018){Maeda}, {Jiang}, {Shigeyama}, \& {Doi}}]{Maeda2018}
{Maeda}, K., {Jiang}, J.-a., {Shigeyama}, T., \& {Doi}, M. 2018, \apj, 861, 78, \dodoi{10.3847/1538-4357/aac8d8}

\bibitem[{{Maeda} {et~al.}(2014){Maeda}, {Kutsuna}, \& {Shigeyama}}]{Maeda2014}
{Maeda}, K., {Kutsuna}, M., \& {Shigeyama}, T. 2014, \apj, 794, 37, \dodoi{10.1088/0004-637X/794/1/37}

\bibitem[{{Magee} \& {Maguire}(2020)}]{Magee2020}
{Magee}, M.~R., \& {Maguire}, K. 2020, \aap, 642, A189, \dodoi{10.1051/0004-6361/202037870}

\bibitem[{{Maoz} {et~al.}(2014){Maoz}, {Mannucci}, \& {Nelemans}}]{Maoz2014}
{Maoz}, D., {Mannucci}, F., \& {Nelemans}, G. 2014, \araa, 52, 107, \dodoi{10.1146/annurev-astro-082812-141031}

\bibitem[{{Matzner} \& {McKee}(1999)}]{Matzner99}
{Matzner}, C.~D., \& {McKee}, C.~F. 1999, \apj, 510, 379, \dodoi{10.1086/306571}

\bibitem[{{Moriya} {et~al.}(2023){Moriya}, {Mazzali}, {Ashall}, \& {Pian}}]{Moriya2023}
{Moriya}, T.~J., {Mazzali}, P.~A., {Ashall}, C., \& {Pian}, E. 2023, \mnras, 522, 6035, \dodoi{10.1093/mnras/stad1386}

\bibitem[{{Nakar}(2020)}]{Nakar2020}
{Nakar}, E. 2020, \physrep, 886, 1, \dodoi{10.1016/j.physrep.2020.08.008}

\bibitem[{{Nakar} \& {Piro}(2014)}]{Nakar2014}
{Nakar}, E., \& {Piro}, A.~L. 2014, \apj, 788, 193, \dodoi{10.1088/0004-637X/788/2/193}

\bibitem[{{Ni} {et~al.}(2024){Ni}, {Moon}, {Drout}, {Lee}, {Sandoval}, {Shin}, {Park}, {Kim}, \& {Oh}}]{Ni2024}
{Ni}, Y.~Q., {Moon}, D.-S., {Drout}, M.~R., {et~al.} 2024, arXiv e-prints, arXiv:2408.06287, \dodoi{10.48550/arXiv.2408.06287}

\bibitem[{{Noebauer} {et~al.}(2017){Noebauer}, {Kromer}, {Taubenberger}, {Baklanov}, {Blinnikov}, {Sorokina}, \& {Hillebrandt}}]{Noebauer2017}
{Noebauer}, U.~M., {Kromer}, M., {Taubenberger}, S., {et~al.} 2017, \mnras, 472, 2787, \dodoi{10.1093/mnras/stx2093}

\bibitem[{{Nomoto}(1982)}]{Nomoto1982}
{Nomoto}, K. 1982, \apj, 257, 780, \dodoi{10.1086/160031}

\bibitem[{{Nugent} {et~al.}(1997){Nugent}, {Baron}, {Branch}, {Fisher}, \& {Hauschildt}}]{Nugent1997}
{Nugent}, P., {Baron}, E., {Branch}, D., {Fisher}, A., \& {Hauschildt}, P.~H. 1997, \apj, 485, 812, \dodoi{10.1086/304459}

\bibitem[{{Pakmor} {et~al.}(2022){Pakmor}, {Callan}, {Collins}, {de Mink}, {Holas}, {Kerzendorf}, {Kromer}, {Neunteufel}, {O'Brien}, {R{\"o}pke}, {Ruiter}, {Seitenzahl}, {Shingles}, {Sim}, \& {Taubenberger}}]{Pakmor22}
{Pakmor}, R., {Callan}, F.~P., {Collins}, C.~E., {et~al.} 2022, \mnras, 517, 5260, \dodoi{10.1093/mnras/stac3107}

\bibitem[{{Papish} {et~al.}(2015){Papish}, {Soker}, {Garc{\'\i}a-Berro}, \& {Aznar-Sigu{\'a}n}}]{Papish15}
{Papish}, O., {Soker}, N., {Garc{\'\i}a-Berro}, E., \& {Aznar-Sigu{\'a}n}, G. 2015, \mnras, 449, 942, \dodoi{10.1093/mnras/stv337}

\bibitem[{{Piro} {et~al.}(2010){Piro}, {Chang}, \& {Weinberg}}]{Piro2010}
{Piro}, A.~L., {Chang}, P., \& {Weinberg}, N.~N. 2010, \apj, 708, 598, \dodoi{10.1088/0004-637X/708/1/598}

\bibitem[{{Piro} \& {Morozova}(2014)}]{Piro2014b}
{Piro}, A.~L., \& {Morozova}, V.~S. 2014, \apjl, 792, L11, \dodoi{10.1088/2041-8205/792/1/L11}

\bibitem[{{Piro} \& {Morozova}(2016)}]{Piro2016}
---. 2016, \apj, 826, 96, \dodoi{10.3847/0004-637X/826/1/96}

\bibitem[{{Piro} \& {Nakar}(2014)}]{Piro2014a}
{Piro}, A.~L., \& {Nakar}, E. 2014, \apj, 784, 85, \dodoi{10.1088/0004-637X/784/1/85}

\bibitem[{{Polin} {et~al.}(2019){Polin}, {Nugent}, \& {Kasen}}]{Polin2019}
{Polin}, A., {Nugent}, P., \& {Kasen}, D. 2019, \apj, 873, 84, \dodoi{10.3847/1538-4357/aafb6a}

\bibitem[{{Sakurai}(1960)}]{Sakuria1960}
{Sakurai}, A. 1960, CPAM, 13, 353, \dodoi{10.1002/cpa.3160130303}

\bibitem[{{Shen} \& {Bildsten}(2014)}]{Shen2014}
{Shen}, K.~J., \& {Bildsten}, L. 2014, \apj, 785, 61, \dodoi{10.1088/0004-637X/785/1/61}

\bibitem[{{Shen} {et~al.}(2021){Shen}, {Blondin}, {Kasen}, {Dessart}, {Townsley}, {Boos}, \& {Hillier}}]{Shen2021}
{Shen}, K.~J., {Blondin}, S., {Kasen}, D., {et~al.} 2021, \apjl, 909, L18, \dodoi{10.3847/2041-8213/abe69b}

\bibitem[{{Shen} {et~al.}(2024){Shen}, {Boos}, \& {Townsley}}]{Shen2024}
{Shen}, K.~J., {Boos}, S.~J., \& {Townsley}, D.~M. 2024, arXiv e-prints, arXiv:2405.19417, \dodoi{10.48550/arXiv.2405.19417}

\bibitem[{{Shen} {et~al.}(2018{\natexlab{a}}){Shen}, {Kasen}, {Miles}, \& {Townsley}}]{Shen2018a}
{Shen}, K.~J., {Kasen}, D., {Miles}, B.~J., \& {Townsley}, D.~M. 2018{\natexlab{a}}, \apj, 854, 52, \dodoi{10.3847/1538-4357/aaa8de}

\bibitem[{{Shen} {et~al.}(2018{\natexlab{b}}){Shen}, {Boubert}, {G{\"a}nsicke}, {Jha}, {Andrews}, {Chomiuk}, {Foley}, {Fraser}, {Gromadzki}, {Guillochon}, {Kotze}, {Maguire}, {Siebert}, {Smith}, {Strader}, {Badenes}, {Kerzendorf}, {Koester}, {Kromer}, {Miles}, {Pakmor}, {Schwab}, {Toloza}, {Toonen}, {Townsley}, \& {Williams}}]{Shen2018}
{Shen}, K.~J., {Boubert}, D., {G{\"a}nsicke}, B.~T., {et~al.} 2018{\natexlab{b}}, \apj, 865, 15, \dodoi{10.3847/1538-4357/aad55b}

\bibitem[{{Shvartzvald} {et~al.}(2024){Shvartzvald}, {Waxman}, {Gal-Yam}, {Ofek}, {Ben-Ami}, {Berge}, {Kowalski}, {B{\"u}hler}, {Worm}, {Rhoads}, {Arcavi}, {Maoz}, {Polishook}, {Stone}, {Trakhtenbrot}, {Ackermann}, {Aharonson}, {Birnholtz}, {Chelouche}, {Guetta}, {Hallakoun}, {Horesh}, {Kushnir}, {Mazeh}, {Nordin}, {Ofir}, {Ohm}, {Parsons}, {Pe'er}, {Perets}, {Perdelwitz}, {Poznanski}, {Sadeh}, {Sagiv}, {Shahaf}, {Soumagnac}, {Tal-Or}, {Santen}, {Zackay}, {Guttman}, {Rekhi}, {Townsend}, {Weinstein}, \& {Wold}}]{Shvartzvald2024}
{Shvartzvald}, Y., {Waxman}, E., {Gal-Yam}, A., {et~al.} 2024, \apj, 964, 74, \dodoi{10.3847/1538-4357/ad2704}

\bibitem[{{Stritzinger} {et~al.}(2018){Stritzinger}, {Shappee}, {Piro}, {Ashall}, {Baron}, {Hoeflich}, {Holmbo}, {Holoien}, {Phillips}, {Burns}, {Contreras}, {Morrell}, \& {Tucker}}]{Stritzinger2018}
{Stritzinger}, M.~D., {Shappee}, B.~J., {Piro}, A.~L., {et~al.} 2018, \apjl, 864, L35, \dodoi{10.3847/2041-8213/aadd46}

\bibitem[{{Taam}(1980)}]{Taam1980}
{Taam}, R.~E. 1980, \apj, 242, 749, \dodoi{10.1086/158509}

\bibitem[{{Tanikawa} {et~al.}(2019){Tanikawa}, {Nomoto}, {Nakasato}, \& {Maeda}}]{Tanikawa19}
{Tanikawa}, A., {Nomoto}, K., {Nakasato}, N., \& {Maeda}, K. 2019, \apj, 885, 103, \dodoi{10.3847/1538-4357/ab46b6}

\bibitem[{{Townsley} {et~al.}(2019){Townsley}, {Miles}, {Shen}, \& {Kasen}}]{Townsley2019}
{Townsley}, D.~M., {Miles}, B.~J., {Shen}, K.~J., \& {Kasen}, D. 2019, \apjl, 878, L38, \dodoi{10.3847/2041-8213/ab27cd}

\bibitem[{{Townsley} {et~al.}(2012){Townsley}, {Moore}, \& {Bildsten}}]{Townsley2012}
{Townsley}, D.~M., {Moore}, K., \& {Bildsten}, L. 2012, \apj, 755, 4, \dodoi{10.1088/0004-637X/755/1/4}

\bibitem[{{Tucker}(2025)}]{Tucker2025}
{Tucker}, M.~A. 2025, \mnras, 538, L1, \dodoi{10.1093/mnrasl/slae121}

\bibitem[{{Werner} {et~al.}(2024){Werner}, {El-Badry}, {G{\"a}nsicke}, \& {Shen}}]{Werner2024}
{Werner}, K., {El-Badry}, K., {G{\"a}nsicke}, B.~T., \& {Shen}, K.~J. 2024, \aap, 689, L6, \dodoi{10.1051/0004-6361/202451635}

\bibitem[{{Woosley} \& {Kasen}(2011)}]{Woosley2011}
{Woosley}, S.~E., \& {Kasen}, D. 2011, \apj, 734, 38, \dodoi{10.1088/0004-637X/734/1/38}

\bibitem[{{Woosley} {et~al.}(1986){Woosley}, {Taam}, \& {Weaver}}]{Woosley1986}
{Woosley}, S.~E., {Taam}, R.~E., \& {Weaver}, T.~A. 1986, \apj, 301, 601, \dodoi{10.1086/163926}

\bibitem[{{Woosley} \& {Weaver}(1994)}]{Woosley1994}
{Woosley}, S.~E., \& {Weaver}, T.~A. 1994, \apj, 423, 371, \dodoi{10.1086/173813}

\bibitem[{{Xi} {et~al.}(2024){Xi}, {Wang}, {Li}, {Liu}, {Yan}, {Lin}, {Zhao}, {Filippenko}, {Zheng}, {Brink}, {Yang}, {Ehgamberdiev}, {Mirzaqulov}, {Reguitti}, {Pastorello}, {Tomasella}, {Cai}, {Zhang}, {Li}, {Zhang}, {Sai}, {Chen}, {Liu}, {Ma}, \& {Xiang}}]{Xi2024}
{Xi}, G., {Wang}, X., {Li}, G., {et~al.} 2024, \mnras, 527, 9957, \dodoi{10.1093/mnras/stad3691}

\bibitem[{{Yuan} {et~al.}(2015){Yuan}, {Zhang}, {Feng}, {Zhang}, {Ling}, {Zhao}, {Deng}, {Qiu}, {Osborne}, {O'Brien}, {Willingale}, {Lapington}, {Fraser}, \& {the Einstein Probe team}}]{Yuan15}
{Yuan}, W., {Zhang}, C., {Feng}, H., {et~al.} 2015, arXiv e-prints, arXiv:1506.07735, \dodoi{10.48550/arXiv.1506.07735}

\end{thebibliography}

\end{document}